\documentclass[useAMS,usenatbib,usegraphicx]{mn2e}
\usepackage{url}
\usepackage{hyperref}
\usepackage{threeparttable}
\usepackage{amssymb}
\usepackage{capt-of}
\def\aj{AJ}%
\def\apj{ApJ}%
\def\apjl{ApJ}%
\def\apjs{ApJS}%
\def\aap{A\&A}%
\def\mnras{MNRAS}%
\def\nat{Nature}%
\def\hi{H\,{\sc i}}
\def\m20{$\rm M_{20}$}
\def\gm{$\rm G_{M}$}

\begin{document}

\title[\hi \ Morphology in Virgo]{Quantified \hi \ Morphology V: \hi \ Disks in the Virgo Cluster}

\author[B.W. Holwerda et al.]{B. W. Holwerda,$^{1,2}$\thanks{E-mail:
benne.holwerda@esa.int} N. Pirzkal,$^{3}$ W.J.G. de Blok,$^{2}$ and W. van Driel$^{4}$\\
$^{1}$ European Space Agency, ESTEC, Keplerlaan 1, 2200 AG, Noordwijk, the Netherlands\\
$^{2}$ Astrophysics, Cosmology and Gravity Centre (ACGC), \\
Astronomy Department, University of Cape Town, Private Bag X3, 7700 Rondebosch, Republic of South Africa\\
$^{3}$ Space Telescope Science Institute, Baltimore, MD 21218, USA\\
$^{4}$ GEPI, Observatoire de Paris, CNRS, Universit\'e Paris Diderot, 5 place Jules Janssen, 92190 Meudon, France
}

\date{Accepted xxxx  Received xxxx; in original form xxxx}

\pagerange{\pageref{firstpage}--\pageref{lastpage}} \pubyear{2011}

\maketitle

\label{firstpage}

\begin{abstract}
We explore the quantified morphology of atomic hydrogen (\hi) disks in the Virgo cluster. These galaxies display a wealth of phenomena in their \hi \ morphology, e.g., tails, truncation and warps. These morphological disturbances are related to the ram-pressure stripping and tidal interaction that galaxies undergo in this dense cluster environment. 
To quantify the morphological transformation of the \hi \ disks, we compute the morphological parameters of Concentration-Asymmetry-Smoothness, Gini and \m20 ~ and our own \gm \ for 51 galaxies in 48  \hi \ column density maps from the VLA Imaging of Virgo spirals in Atomic gas (VIVA) project.

Some morphological phenomena can be identified in this space of relatively low resolution \hi \ data. 
Truncation of the \hi \ disk can be cleanly identified via the Concentration parameter (C$<$1) and Concentration can also be used to identify \hi \ deficient disks (1$<$C$<$5). 
Tidal interaction is typically identified using combinations of these morphological parameters, applied to (optical) images of galaxies. 
We find that some selection criteria (Gini-$M_{20}$, Asymmetry, and a modified Concentration-$M_{20}$) are still applicable for the coarse ($\sim$15$"$ FWHM) VIVA \hi \ data. 
We note that Asymmetry is strongly affected by the choice for the center of these galaxies. The phenomena of tidal tails can be reasonably well identified using the Gini-\m20 ~ criterion (60 \% of galaxies with tails identified but with as many contaminants).

Ram-pressure does move \hi \ disks into and out of most of our interaction criteria: the ram-pressure sequence identified by Vollmer et al. (2009) tracks into and out of some of these criteria (Asymmetry based and the Gini-$M_{20}$ selections, but not the Concentration-\m20 or the \gm based ones). Therefore, future searches for interaction using \hi \ morphologies should take ram-pressure into account as a mechanism to disturb \hi \ disks enough to make them appear as gravitationally interacting. One mechanism would be to remove all the \hi \ deficient (C$<$5) disks from the sample, as these have undergone more than one \hi \ removal mechanism. 

\end{abstract}


\section{Introduction}
\label{s:intro}

The Virgo cluster represents an ideal laboratory for the study of galaxy evolution in a cluster environment. It is one of the nearest galaxy clusters (D$\sim$17 Mpc),  a relatively populous system (Abell richness class I), consisting of some two thousand catalogued members in the Virgo Cluster Catalogue \citep[VCC,][]{Binggeli85}. The Virgo cluster is spiral-rich and dynamically young, i.e., made of galaxy subgroups which are falling into the main cluster. The most prominent groups are those associated with M86 to the west and M49 to the south.  A particularity of the Virgo cluster is that the distribution of its hot intracluster medium is strongly peaked on the central cD galaxy M87 \citep{Bohringer94}, and overall highly sub-structured. The Virgo cluster is therefore ideal to study the effects of both gravitational interaction and interaction with the intracluster medium on the atomic gas (\hi) disks of spiral members. 

Morphological signs of tidal interaction are usually detected in the restframe ultraviolet or blue-filter optical images of the stellar disk. A series of morphological parameters, originally developed for galaxy classification, have been used extensively in recent years to identify disturbed galaxies, and to infer the merger fractions and rates over cosmological time scales \citep{Lotz04, Lotz08b, CAS, Conselice03b, Conselice08b, Conselice09b, Bundy05, Yan05, Ravindranath06, Trujillo07, Jogee09}. The observational benefits are that the Hubble Space Telescope observations of high-redshift galaxies and ground-based observations of nearby galaxies (e.g., SDSS) probe the same wavelength, physical structure and volumes. However, in the nearby Universe, the 21 cm perspective on galaxies is set to catch up to the optical surveys in terms of numbers of galaxies studied. Currently under construction are two new radio observatories: the South African Karoo Array Telescope \citep[MeerKAT;][]{meerkat1, MeerKAT,meerkat2}, and the Australian SKA Pathfinder \citep[ASKAP;][]{askap2, askap1, ASKAP, askap3,askap4}. Together with refurbished observatories such as the Extended Very Large Array \citep[EVLA;][]{evla} and the APERture Tile In Focus instrument \citep[APERTIF;][]{apertif,apertif2} on the Westerbork Synthesis Radio Telescope (WSRT), detailed \hi \ maps of thousands of galaxies will become available. The unprecedented volume of data implies that the data reduction and subsequent  morphological classification can only be done through automated procedures.

In the previous papers in this series \citep{Holwerdapra09, Holwerda10b, Holwerda10c, Holwerda10d, Holwerda10e}, we have shown that the morphological classifiers that are in use to identify mergers in the optical can be extremely useful for the \hi \ perspective. The \hi \ observations are not as high-resolution as the optical disk but this is compensated for by the larger extent of the atomic gas disk and the relative sensitivity of gas to an interaction -- for which much anecdotal evidence already existed in ``The \hi \ Rogues Galaxies" catalogue \citep{Hibbard01}\footnote{\url{http://www.nrao.edu/astrores/HIrogues/RoguesLiving.shtml}}. 
The sensitivity of \hi \ surveys to gas-rich and minor mergers will provide the calibration for higher redshifts where these types of mergers are expected to be the dominant type of galaxy interaction \citep{Lotz10a,Lotz10b}. The Square Kilometre Array \citep{ska} will subsequently be able to resolve \hi \ disks at these higher redshifts.

The \hi \ disks of spiral galaxies are also extremely sensitive to ram pressure by the intergalactic or intracluster medium. The Virgo cluster is an ideal laboratory to test the effects of both gravitational interaction and intracluster medium stripping on the quantified morphology of \hi \ disks. There are detailed models of the gas stripping sequence \citep{Vollmer09}, and a uniform \hi \ data set from the VLA Imaging of Virgo spirals in Atomic gas (VIVA) project \citep{Chung09}, with detailed notes on individual galaxies identifying morphological phenomena.

In this paper, we explore the distribution of the \hi \ morphological parameters from the VIVA column density maps in relation to estimates by \cite{Chung07},  \cite{Chung09} or \cite{Vollmer09} of the star-formation rate, \hi \ deficiency, stripping and gravitational harassment. First we discuss data and morphological parameters (\S \ref{s:data} and \ref{s:param}), but readers already familiar with these can skip to results and conclusions (\S \ref{s:results} and \ref{s:concl}).

\section{VIVA Data}
\label{s:data}

The VIVA project \citep{Chung09} has compiled the largest high-resolution imaging \hi \ database of the Virgo cluster to date. They have published 48 VLA \hi \ data cubes and maps with 51 Virgo cluster spiral galaxies. These are some 40 new observations with the remainder from reprocessed archival data. We obtained the zero order moment maps from \url{http://www.astro.yale.edu/viva/} and converted to \hi \ column density maps using the expressions in \cite{Walter08}. 
The spatial resolution of VIVA data is approximately 15$"$ (or $\sim$1.2 kpc at the distance of Virgo), and the typical sensitivity is $\sim3-5 \times 10^{19}$  cm$^{-2}$ in 3$\sigma$ per 10 km/s channel. 
VIVA data show that (i) the galaxies located in the core of the cluster predominantly have truncated \hi \  disks with some gas displaced from  the galactic disk and (ii) some 7 out of the 50 sample galaxies ($\sim$10\%) show long, one-sided \hi \ tails \citep{Chung07}. These tails occur in galaxies located at intermediate projected distances between 0.5 and 1 Mpc from the cluster centre. 
From the sample listed in Table 1 of \cite{Chung09}, we did not use IC 3418 as it is a non-detection in the VIVA survey and Holmberg VII because it does not have enough continuous \hi \ flux to meaningfully compute morphological parameters.


\section{Morphological Parameters}
\label{s:param}

Two morphological parameter schemes are now in wide use: Concentration-Asymmetry-Smoothness \citep[CAS,][]{CAS} and Gini--\m20 ~ \citep{Lotz04}. 
In an image with $n$ pixels with intensities $I(i,j)$ at pixel position $(i,j)$, CAS the morphological parameters are defined as:
\begin{equation}
C = 5 ~ log \left( {r_{80} \over  r_{20} } \right)
\label{eq:c}
\end{equation}
\begin{equation}
A = {\Sigma_{i,j} | I(i,j) - I_{180}(i,j) |  \over \Sigma_{i,j} | I(i,j) |  }
\label{eq:a}
\end{equation}
\begin{equation}
S = {\Sigma_{i,j} | I(i,j) - I_{S}(i,j) | \over \Sigma_{i,j} | I(i,j) | }
\label{eq:s}
\end{equation}
\noindent where $r_{\%}$ is the radius which includes that percentage of the intensity of the object, $I_{180}(i,j)$ is the value of the pixel in the rotated image and $I_{S}(i,j)$ is the same pixel in the image after smoothing. 

Gini and \m20 ~ are defined as:
\begin{equation}
G = {1\over \bar{I} n (n-1)} \Sigma_i (2i - n - 1) | I_i |
\label{eq:g}
\end{equation}
\begin{equation}
M_{20} = \log \left( {\Sigma_i^k M_i  \over  M_{tot}}\right), ~ {\rm for ~ which} ~ \Sigma_i^k I_i < 0.2 ~ I_{tot} {\rm ~ is ~ true}.\\
\label{eq:m20}
\end{equation}
\noindent with $M_{tot} = \Sigma M_i = \Sigma I_i [(x_i - x_c)^2 + (y_i - y_c)^2]$, where ($x_c$,$y_c$) is the galaxy's central position, and $I_i$ is the intensity of pixel $i$ in an flux-ordered list. Pixel $k$ is the pixel marking the 20\% point in this list. $\bar{I}$ is the mean pixel value. 
The Gini parameter is an indicator of inequality in the distribution of pixel values; Gini=1 is perfect inequality with one pixel containing all the flux from an object and Gini=0 is perfect equality with each pixel containing an equal fraction of the flux. Values in between indicate how concentrated the flux from this object is in a few bright areas. The \m20 ~ parameter is an indication whether or not the brightest parts of the object are close to the center or spread throughout the object. 

In addition to these well-established parameters, we compute the Gini coefficient of the second order moment $I_i \times [ (x_i - x_c)^2 + (y_i - y_c)^2]$, which we called $G_M$ \citep{Holwerda10c}. For all these parameters, we compute the uncertainty by varying the input centre the resolution and randomizing the pixel values. The exception is the Gini parameter as this does not rely on the centre as input. For the Gini error, we jacknife the pixel selection.
For a more detailed description of these parameters, we refer the reader to the previous papers in this series \citep{Holwerda10b,Holwerda10c} and the original papers introducing these \citep{CAS,Lotz04}. We compute these parameters for the area defined by the contour level of $1 \times 10^{19}$  cm$^{-2}$ (the $1\sigma$ level).

In \cite{Holwerda10c}, we identified several parts of parameter space where tidally interacting galaxies reside for the WHISP sample. 
To identify the interacting galaxies, we used a subsample of 154 spirals and dwarf galaxies in WHISP, which have independent visual estimates of interaction
from either \cite{Swaters02} or \cite{Noordermeer05}. 
The criteria from the literature and our previous paper are identified in Figures \ref{f:tail} and \ref{f:int} with dashed and dotted lines respectively. Criteria from the literature are:
\begin{equation}
A> 0.38 
\label{eq:lcrit1}
\end{equation}
\noindent from \cite{CAS}, marked by the dashed lines in Figures \ref{f:tail} and \ref{f:int}, panels IV, V and VI. 
\begin{equation}
G > -0.115 \times M_{20} + 0.384
\label{eq:lcrit2}
\end{equation}
\noindent from \cite{Lotz04}, marked by the dashed lines in panel II of Figures \ref{f:tail} and \ref{f:int}.
\begin{equation}
G > -0.4 \times A + 0.66 ~ \rm or ~ A > 0.4
\label{eq:lcrit3}
\end{equation}
\noindent \cite{Lotz10a}, marked by the dashed lines in panel IV of Figures \ref{f:tail} and \ref{f:int}.

We defined three criteria based on the WHISP \hi \ morphologies:
\begin{equation}
G_M > 0.6,
\label{eq:crit1}
\end{equation}
\noindent makred by the vertical dotted line in Figures \ref{f:tail} and \ref{f:int}, panels I, III, VI and X,
\begin{equation}
A > -0.2 \times M_{20} + 0.25,
\label{eq:crit2}
\end{equation}
\noindent marked by the dotted line in Figures \ref{f:tail} and \ref{f:int}, panel V, and, 
\begin{equation}
C > -5 \times M_{20} + 3, 
\label{eq:crit3}
\end{equation}
\noindent which is the dotted lines in Figures \ref{f:tail} and \ref{f:int}, panel IX.

These criteria select a merger from its \hi \ morphology for different time scales. In \cite{Holwerda10d}, we quantified the typical time scale for 
a merger to be selected by each of these criteria using \hi \  maps of merger simulations \citep[originally presented in][]{Cox06a,Cox06b}. These can be compared to merger simulations of an isolated disk, evolving passively. The comparison shows how long a merger is selected as well as the level of contamination by passive disks.
At the resolution of the WHISP sample, interactions were identifiable as such by their \hi \ morphology for about a Gyr. This is a similar time scale as the one for optical morphological identifications and better than the close pair method.
Our goal is to explore if the above criteria could be applied to \hi \ data in a denser environment, where many other phenomena play a role in the \hi \ morphology as well as to explore how many of those other phenomena leave a signature in the \hi \ morphology.


\section{Results: \hi \ Morphology}
\label{s:results}
 
\begin{figure*}
\begin{center}
\includegraphics[width=1.0\textwidth]{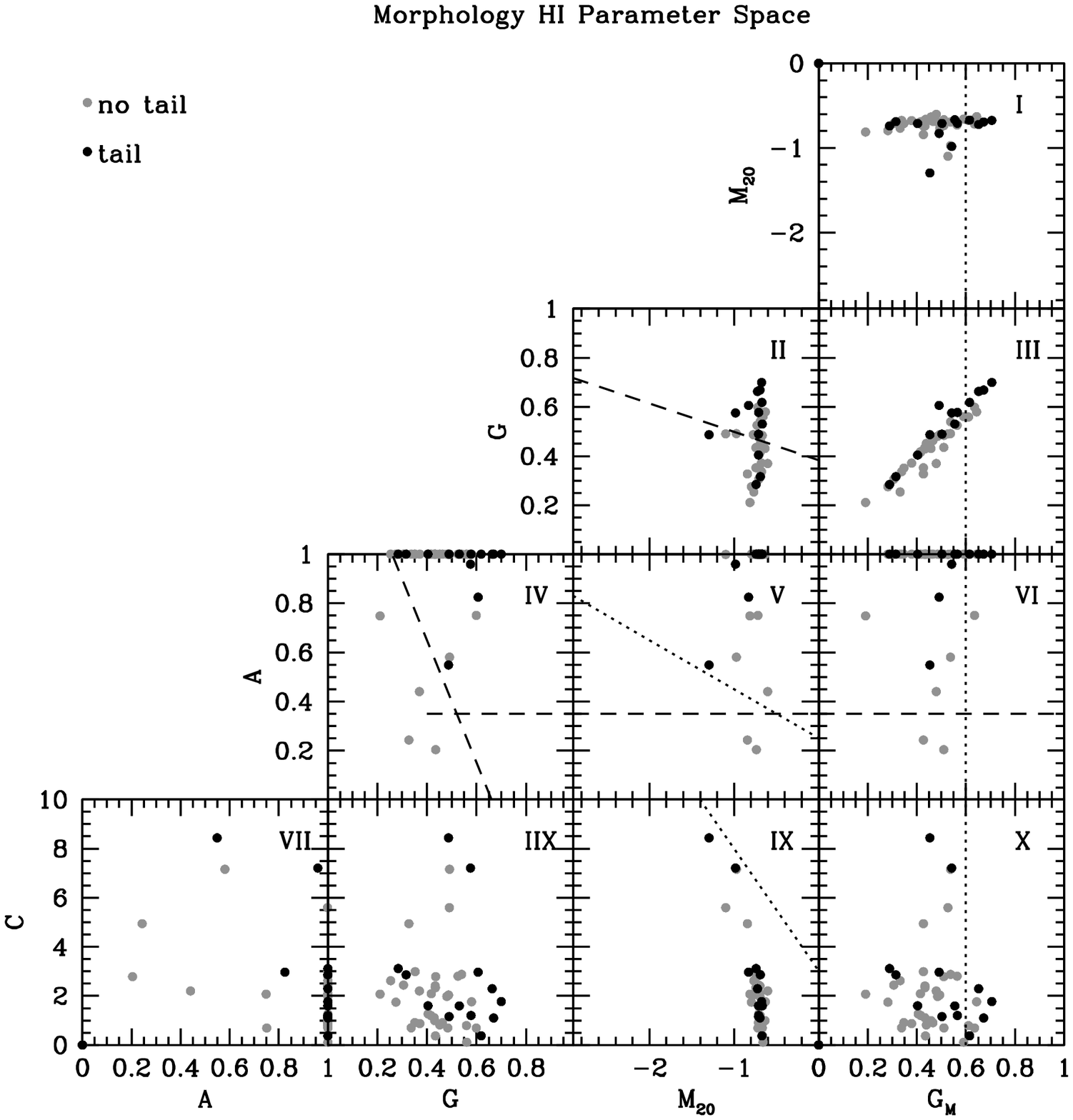}
\caption{The spread of morphological parameters. Galaxies with an \protect\hi \ tail, as identified in \protect\cite{Chung07} and \protect\cite{Chung09} are marked with solid black points. 
The (optical) merger selection criteria from the literature are marked with dashed lines in panel II (equation \protect\ref{eq:lcrit2}), panel IV (equations \protect\ref{eq:lcrit1} and \ref{eq:lcrit3}), and V and VI (equation \protect\ref{eq:lcrit1}). Our selection criteria from \protect\cite{Holwerda10c} are marked with dotted lines; the $G_M$ criterion in panels I, III, VI and X (equation \protect\ref{eq:crit1}), the $A$-\m20 criterion in panel V (\protect\ref{eq:crit2}) and the C-\m20 criterion in panel IX (equation \protect\ref{eq:crit3}). }
\label{f:tail}
\end{center}
\end{figure*}

Figures \ref{f:tail} and \ref{f:int} show the spread of morphological parameters for the entire VIVA sample with different morphological classifications from \cite{Chung07,Chung09}. Table \ref{t:viva} lists all the values. Due to the combination of distance and resolution, the VIVA sample has the coarsest physical scales we have applied these morphological parameters to date. In \cite{Holwerda10b} and \cite{Holwerda10d} we showed that these parameters are relatively invariant with distances out to D$\sim$20 Mpc and a resolution of 6", so the VIVA sample is effectively at the limit of this technique with current observatories. 

We used the optical positions (J2000) of the galaxies reported in \cite{Chung09} as the centre of the galaxy ($x_c$,$y_c$) to compute the parameters that depend on a central position (all except the Gini paramter). Our motivations are that the stellar disk marks the centre of the gravitational potential of a galaxy and kinematic and stellar centres generally align well \citep[see][]{THINGS, Trachternach08}. Optical positions of the galaxies are a likely starting point for many of the future \hi \ surveys and \cite{Chung09} reported many tidal \hi \ tails and displacements, which could be quantified using our morphological parameters. Thus, it is entirely possible that some of morphological values (e.g., Asymmetry) are extreme since the centre is not the barycentre of the flux, as is common for optical morphological measurements.

\begin{figure*}
\begin{center}
\includegraphics[width=0.49\textwidth]{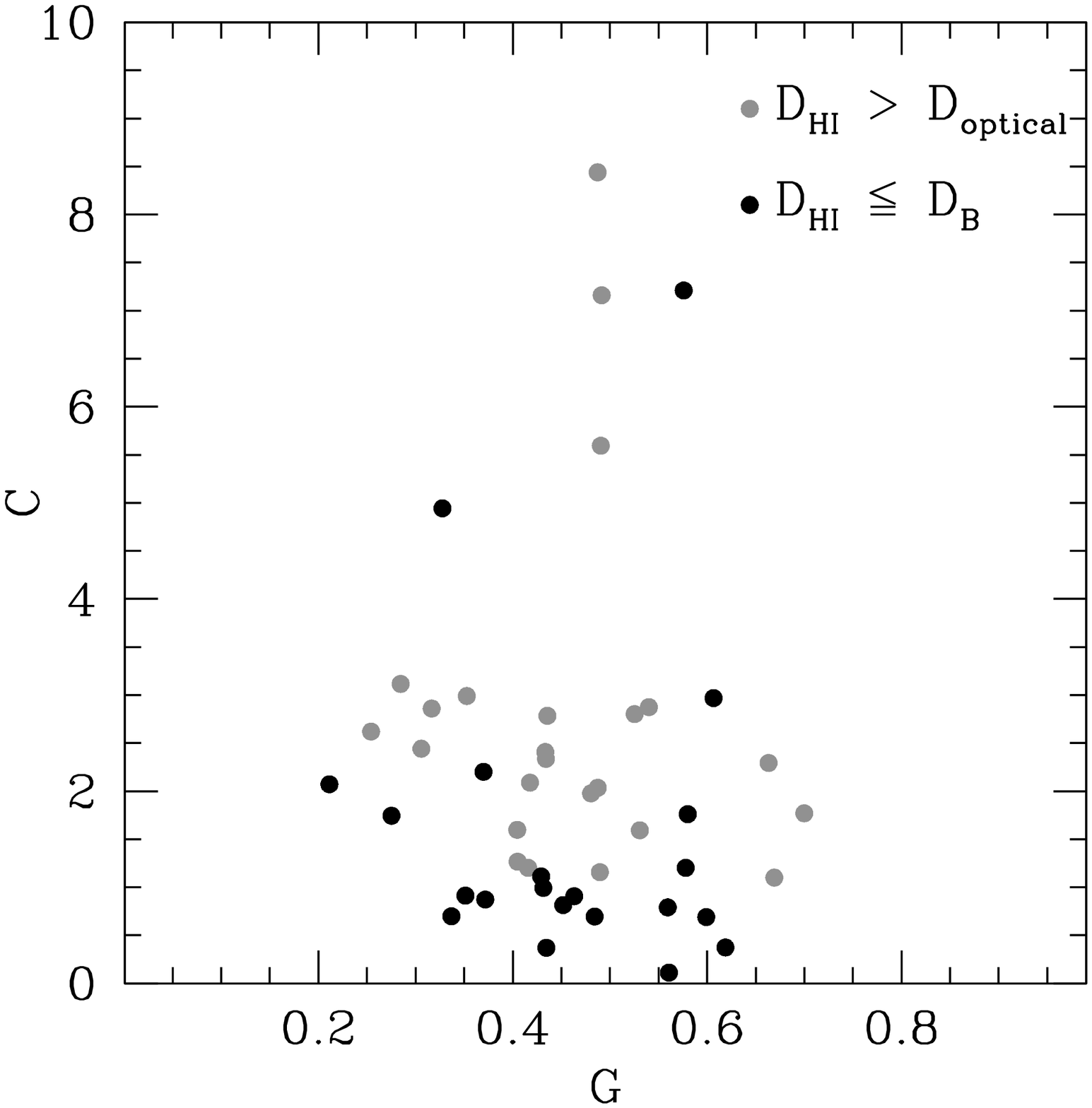}
\includegraphics[width=0.49\textwidth]{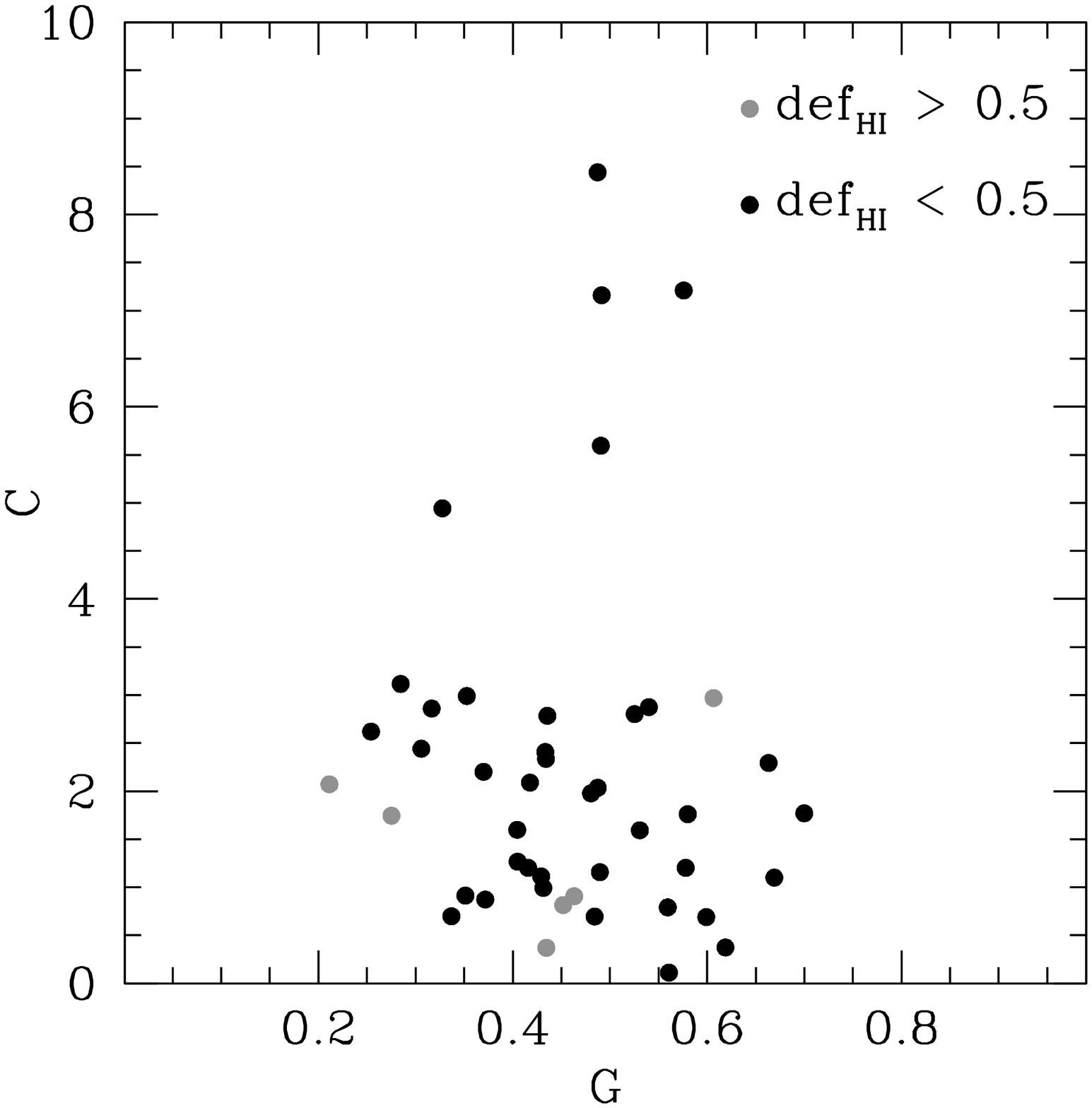}
\caption{The relation between Gini and Concentration for \hi \ truncated galaxies ($D_{HI} \leq D_{B}$, left panel) and \hi \ deficient galaxies (right panel). 
Most of the VIVA sample disks are \hi \ deficient ($ def_{HI} ~ < ~ 1$) and we mark those that are extremely so ($ def_{HI} ~ < ~ 0.5$); galaxies with half a dex or more below the typical \hi \ surface density for their type of spiral \protect\citep{Haynes84}. Typical values for concentration are 5 and above for \hi \ disks in WHISP \protect\citep{Holwerda10c,Holwerda10e}. Therefore the majority of the \hi \ disks in VIVA are more than typically centrally concentrated; a low concentration value (C$<5$) is a good sign of \hi \ deficiency and, in addition, an extremely low concentration value (C$<$1) identifies half of the truncated \hi \ disks.} 
\label{f:hidef}
\end{center}
\end{figure*}

\cite{Chung09} supply extensive notes on the \hi \ morphology and the nature of the \hi \ stripping of individual galaxies. We have summarized these evaluations in Table \ref{t:chung}. \cite{Chung09} note which galaxies are currently undergoing or have recently undergone stripping, as well as the likely origin of this stripping; ram pressure by the intracluster medium or tidal interaction with a close companion.
In addition, they report a parameter for \hi \ deficiency ($def_{HI}$; the galaxy's \hi \ surface density compared to the value typical for spirals), a measure of \hi \ truncation ($\rm D_{HI} / D_B $; the ratio of the \hi \ diameter over the optical, B-band one), and additional notes for indicators of warps in the \hi \ morphology and velocity maps. 
Finally, \cite{Chung09} also list the star-formation properties of their sample, as catalogued and classified by \cite{Koopmann04a, Koopmann04b}; the level of activity and whether it was recently cut-off (truncated star-formation). We have plot the spread of our morphological parameters with each of these estimates marked to explore if there are specific parts of parameter space that single out an \hi \ disk characteristic, similar to the separation of interaction and non-interacting galaxies we found in the WHISP database. 
Figures \ref{f:hidef}, \ref{f:tail}, and \ref{f:int}, show the spread in parameters each with certain phenomena marked, so that they can be identified in the morphological parameter space. We discuss several \hi \ morphological phenomena noted by Chung et al. below.

\subsection{Truncation and Deficiency}
\label{s:hidef}

Figure \ref{f:hidef} shows Gini versus Concentration for \hi \ deficient and truncated \hi \ disks. Both parameters are sensitive to how evenly the flux is distributed over the area of the disk and therefore could be sensitive to both these phenomena.
To identify \hi \ truncation, one could use the Concentration of the \hi \ emission. We have found that values around or below C=1 seem to be very indicative of truncation; the radii containing 20 and 80 percent of the \hi \ flux are very close together (Fig. \ref{f:hidef}, left panel). Given that typical values for concentration in \hi \ are much higher \citep[typically C$\ge$5,][]{Holwerda10c,Holwerda10f}, this may be a good tracer of truncation in large \hi \ surveys.  A value of Concentration below unity selects half of the truncated galaxies in the VIVA sample very cleanly (12 out of 24 truncated disks with no contamination from non-truncated disks). 

The \hi \ deficiency is defined by the surface density compared to the average surface density for spirals \citep[see equation 5 in][]{Chung09}. The majority of VIVA galaxies are \hi \ deficient (44 galaxies with $def_{HI} < 1$). The low values of concentration ($C<5$) appear to be indicative of this \hi \ deficiency. In VIVA, those galaxies with extreme \hi \ deficiency ($def_{HI} < 0.5$) are not much more concentrated than moderately deficient ones (Figure \ref{f:hidef}, right panel), at least at this resolution.
We note that a C$\leq$5 criterion will select some 80 \% of the \hi \  deficient galaxies in the VIVA sample but their number is contaminated (17\% of this selection is not \hi \ deficient). In the upcoming large surveys this makes Concentration a quick assessor on whether or not a spiral has a typical \hi \ content or not.

\subsection{Interaction; \hi \ Tails, and Ram-pressure}
\label{s:inter}

Nearly all the spirals in the Virgo cluster are undergoing, or have recently undergone, some type of interaction. Discerning between the dominant mechanism, tidal or ram-pressure, was a major focus of the work by \cite{Chung09} and \cite{Vollmer09}. Several interaction related morphological phenomena (e.g., warps and \hi \ tails) are noted by \cite{Chung07}.

Table \ref{t:chung} lists many galaxies with \hi \ tails in the VIVA data and several with \hi \ warps. 
One would expect that high values for the Asymmetry would identify \hi \ tails (Figure \ref{f:tail}). However, because we used the optical centres listed by \cite{Chung09}, Asymmetry is already at the maximum value for most of the VIVA galaxies (Figure \ref{f:tail}, panels IV-VII). We note however that the Asymmetry may still be a good indicator of (tidal) \hi \ tails, but in less extreme or possibly better resolved cases.
The \m20-Gini selection criterion (see eq. \ref{eq:lcrit2}) does reasonably separate those galaxies with \hi \ tails from those which do not; 60 \% of the \hi \ tails identified by \cite{Chung07,Chung09} are selected by this criterion (Figure \ref{f:tail}, panel II). However, these are about half the objects selected do not have an \hi \ tail. Therefore,
the Gini-\m20 ~ criterion can be used a likelyhood for an \hi \ tail but it will suffer from contamination.
Warps were identified predominantly by their kinematic signature in \cite{Chung09} and no morphological parameter or combination seems to be able to identify them in the VIVA data.

\begin{figure*}
\begin{center}
\includegraphics[width=\textwidth]{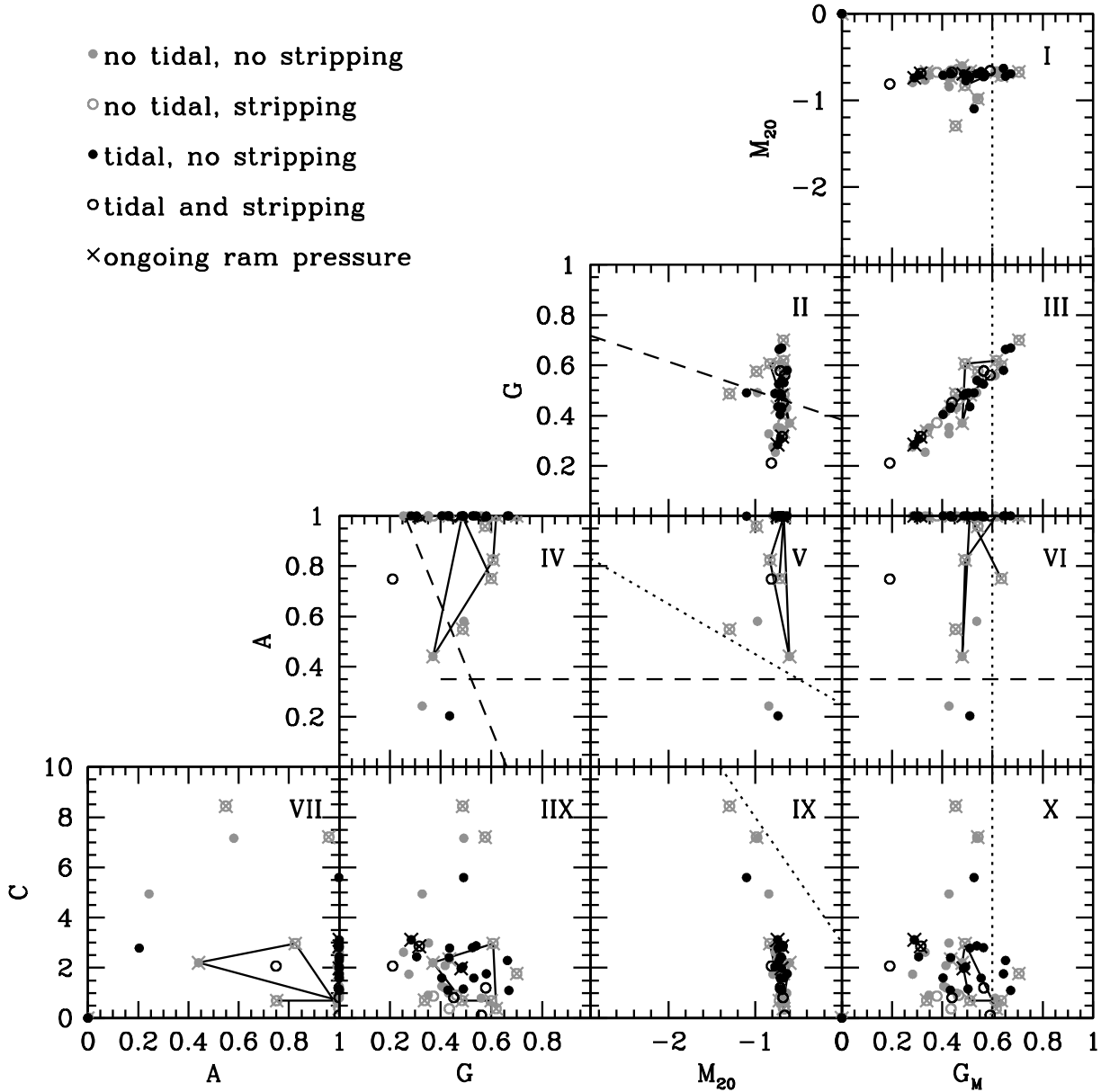}
\caption{The spread of morphological parameters for galaxies undergoing tidal interaction (black symbols), \hi \ stripping (open circles) and ongoing ram-pressure (marked with an x), according to \protect\cite{Chung09}, see Table \ref{t:chung}. 
The (optical) merger selection criteria from the literature are marked with dashed lines in panel II (equation \protect\ref{eq:lcrit2}), panel IV (equations \protect\ref{eq:lcrit1} and \ref{eq:lcrit3}), and V and VI (equation \protect\ref{eq:lcrit1}). Our selection criteria from \protect\cite{Holwerda10c} are marked with dotted lines; the $G_M$ criterion in panels I, III, VI and X (equation \protect\ref{eq:crit1}), the $A$-\m20 criterion in panel V (\protect\ref{eq:crit2}) and the C-\m20 criterion in panel IX (equation \protect\ref{eq:crit3}). 
The solid lines connect the ram-pressure sequence determined by  \protect\cite{Vollmer09} for a select few Virgo cluster galaxies. }
\label{f:int}
\end{center}
\end{figure*}

Figure \ref{f:int} shows the same spread of morphological parameters as Figure \ref{f:tail} with the estimate we gleaned from \cite{Chung09} whether or not the galaxy is stripped, undergoing ram-pressure, or undergoing tidal interaction. The sequence of stripping in the Virgo cluster according to \cite{Vollmer09} is indicated by the solid lines. 
This sequence of six Virgo galaxies does trace a track into and out of certain interaction criteria defined previously in the literature and the third paper in this series.

The extreme values for Asymmetry drive almost all of the VIVA galaxies into the Asymmetry-based merger selection space; the Asymmetry criterion from \cite{CAS}, the Asymmetry-Gini criterion from \cite{Lotz10a} and our Asymmetry-\m20 criterion from \cite{Holwerda10c} (equations \ref{eq:lcrit1}, \ref{eq:lcrit3}, and \ref{eq:crit2} respectively). 
The extreme values for Concentration --linked to \hi \ deficiency and truncation-- similarly make the Concentration--\m20 selection criterion \citep[equation \ref{eq:crit3},][]{Holwerda10c} not applicable for merger selection.
Only the Gini-\m20 criterion (eq. \ref{eq:lcrit2}) appears to separate objects with \hi \ tails moderately as discussed above.

The \gm \ criterion (eq. \ref{eq:crit1}) is not contaminated by galaxies undergoing ram-pressure stripping but does not select tidal interactions well anymore either, something that we attribute to the native VIVA resolution \citep[no visibility times found in ][for the VIVA resolution]{Holwerda10d}. 
Because stripping, \hi \ deficiency and truncation are interlinked, there is the possibility that the Concentration selection criteria also will select those galaxies that are or have been stripped. 


Two effects move \hi \ morphologies into the selection criteria: poorer resolution and additional effects such as the ram-pressure stripping. 
Given how the Vollmer sequence tracks into (and out again) of the merger selection criteria, we can conclude that galaxies undergoing ram-pressure stripping will sometimes be selected as interacting in a morphological selection in future \hi \ surveys, at least those with spatial resolutions similar to VIVA. The addition of some galaxies that are undergoing stripping would, in part, explain the high merger fraction and rate we found in the WHISP survey \citep{Holwerda10e}. 
In \cite{Holwerda10d}, we noted that the lower resolution data of the VIVA sample would have some problems selecting interacting systems from isolated ones, purely on the basis of their \hi \ morphology (Table 8 in the appendix). However, the success with which tidal \hi \ tails could be separated is encouraging. 

Therefore, in future medium-resolution \hi \ surveys (e.g., WALLABY, Koribalski et al, {\em in preparation}), the selection of \hi \ disks with disturbed morphology can be used as a first pass in a search for stripping or interacting galaxies and a merger fraction and rate will have to be corrected for the contamination by stripping galaxies. For example, those galaxies that are \hi \ deficient can be removed before a merger rate is determined but that would bias against denser cluster environments (which do include more other interactions like ram-pressure stripping).
A higher resolution survey of the Virgo cluster galaxies would resolve the population of tidally interacting galaxies.
The separation of ram-pressure stripped galaxies from tidal interacting ones may then be possible without the removal of \hi \ deficient ones.


\subsection{Star Formation}
\label{s:sf}


We explored the link between the morphological parameter space of \hi \ and the star-formation and truncation of star-formation as listed in Table 3 in \cite{Chung09}. Relations are weak as the \hi \ gas is undergoing several other effects other than feedback from star-formation but it is interesting to note that those galaxies with little star-formation also have lower values of Gini for \hi \ (i.e., they are more homogeneous).

\section{Concluding Remarks}
\label{s:concl}

In this paper, we present the morphological parameters commonly used in the optical, applied to the \hi \ column density maps of Virgo Cluster galaxies obtained by the VIVA program. We find that the resolution of these observations ($\sim$15" or 1 kpc) is on the coarse side for this type of analysis but some \hi \  morphological phenomena are identifiable, though with limitations. 
A critical choice is whether or not to use the optical or the \hi \ centres for these galaxies, one that every \hi \ survey will have to make. The choice of centre affects measures such as concentration, \m20, \gm, and especially Asymmetry. Alternatively, one could use the barycentre of the \hi \ emission itself but the optical centre is most likely the centre of the dark matter halo of the galaxy and this approach would likely miss \hi \ displacements like tidal tails. 
We find that of the \hi \ morphological phenomena, \hi \ deficiency and truncation can be identified easily by a low value of the concentration index (Figure \ref{f:hidef}). Warps could not be identified using the morphology parameters and \hi \ tails can be identified moderately well with the combined Gini and \m20 interaction criterion (eq. \ref{eq:lcrit2} \& Figure \ref{f:tail}, panel II) from \cite{Lotz04, Lotz08b}.

Galaxies that are interacting are still identified by some of the morphological criteria from the literature \citep{CAS,Lotz04} and those identified by us specifically for \hi \ in \cite{Holwerda10c} (eq. \ref{eq:lcrit1} -- \ref{eq:crit3}). However, galaxies affected by ram-pressure stripping are ofttimes also selected by these criteria. For example, we note that the ram pressure sequence identified by \cite{Vollmer09} tracks into the interaction part of the parameter space for many of the morphological selection criteria (solid tracks in Figure \ref{f:int}).
Discerning between the origin of the \hi \ stripping --ram pressure or tidal-- is, however, not possible with these morphological parameters, at least at VIVA's spatial resolution.  Thus any selection of interacting systems in a survey based purely on the \hi \ morphology parameters will include those that are undergoing ram-pressure stripping in a denser environment. A straightforward correction would be to exclude all \hi \ deficient ($C<5$) galaxies from a morphological selection of interacting galaxies.

\section*{Acknowledgments}

We wish to thank the anonymous referee for his or her comments, helping us to greatly improve the final manuscript. We thank the VIVA collaboration for making their data public and easily accessible. The National Radio Astronomy Observatory is a facility of the National Science Foundation operated under cooperative agreement by Associated Universities, Inc. We acknowledge support from the National Research Foundation of South Africa. The work of B.W. Holwerda  and W.J.G. de Blok is based upon research supported by the South African Research Chairs Initiative of the Department of Science and Technology and the National Research Foundation.


\begin{threeparttable}[f]
\begin{tabular}{l l l l l l l l}
NGC		&	VCC		& strip	& ram	&tidal	& warp	& tail & note\\ 
\hline
\hline
 4064	 & \dots	& \checkmark	 &  	 & \checkmark	 &  	 &  	& \\
 4189	 & 89		&  	 &  	 &  	 &  	 &  	& \\
 4192	 & 92		&  	 &  	 &  	 & \checkmark	 &  	& \\
 4216	 & 167	&  	 &  	 &  	 &  	 &  	& \\
 4222	 & 187	&  	 &  	 &  	 & \checkmark	 &  	& \\
 4254	 & 307	 &  	 &  	 & \checkmark	 &  	 & \checkmark	& \\
 4293	 & 460	 & \checkmark	 &  	 & \checkmark	 &  	 &  	& \\
 4294	 & 465	 &  	 &  	 & \checkmark	 &  	 & \checkmark	& a \\
 4298	 & 483	 &  	 &  	 & \checkmark	 &  	 &  	& \\
 4299	 & 491	 & \checkmark	 & \checkmark	 & \checkmark	 &  	& a \\
 4302	 & 497	 & \checkmark	 & \checkmark	 &  	 &  	 & \checkmark	& a\\
 4321	 & 596	 &  	 &  	 & \checkmark	 &  	 &  	& \\
 4330	 & 630	 & \checkmark	 & \checkmark	 &  	 &  	 & \checkmark	& a, I\\
 4351	 & 692	 &  	 & \checkmark	 & \checkmark	 &  	 &  	& \\
 4380	 & 792	 &  	 &  	 &  	 &  	 &  	& \\
 4383	 & 801	 &  	 &  	 & \checkmark	 & \checkmark	 &  	& \\
 4388	 & 836	 & \checkmark	 & \checkmark	 &  	 &  	 & \checkmark	& II\\
 4394	 & 857	 &  	 &  	 &  	 &  	 &  	& \\
 4396	 & 865	 & \checkmark	 & \checkmark	 &  	 &  	 & \checkmark	& a\\
 4405	 & 874	 & \checkmark	 &  	 &  	 &  	 &  	& \\
 4402	 & 873	 & \checkmark	 & \checkmark	 &  	 &  	 & \checkmark	& \\
IC3355	 & 945	 &  	 &  	 & \checkmark	 &  	 &  	& \\
 4419	 & 958	 & \checkmark	 &  	 &  	 &  	 &  	& \\
 4424	 & 979	 & \checkmark	 &  	 & \checkmark	 &  	 & \checkmark	& a\\
 4450	 & 1110	 &  	 &  	 &  	 &  	 &  	& \\
IC3392	 & 1126	 & \checkmark	 &  	 &  	 &  	 &  	& \\
 4457	 & 1145	 &  	 &  	 &  	 &  	 &  	& \\
 4501	 & 1401	 &  	 & \checkmark	 &  	 &  	 &  	& III\\
 4522	 & 1516	 & \checkmark	 & \checkmark	 &  	 &  	 &  	& IV\\
 4532	 & 1554	 &  	 &  	 & \checkmark	 & \checkmark	 & \checkmark	& \\
 4535	 & 1555	 &  	 & \checkmark	 &  	 &  	 &  	& \\
 4533	 & 1557	 &  	 &  	 & \checkmark	 &  	 & \checkmark	& \\
 4536	 & 1562	 &  	 &  	 & \checkmark	 &  	 &  	& \\
 4548	 & 1615	 &  	 &  	 &  	 &  	 &  	& \\
 4561	 & \dots	 &  	 &  	 & \checkmark	 &  	 &  	& \\
 4567	 & 1673	 &  	 &  	 & \checkmark	 &  	 & \checkmark	& \\
 4568	 & 1676	 &  	 &  	 & \checkmark	 & \checkmark	 &  	& \\
 4569	 & 1690	 & \checkmark	 & \checkmark	 &  	 &  	 &  	& V\\
 4579	 & 1727	 &  	 &  	 &  	 &  	 &  	& \\
 4580	 & 1730	 & \checkmark	 & \checkmark	 &  	 &  	 &  	& \\
 4606	 & 1859	 & \checkmark	 &  	 & \checkmark	 &  	 &  	& \\
 4607	 & 1868	 &  	 &  	 &  	 &  	 &  	& \\
 4651	 & \dots	 &  	 &  	 & \checkmark	 & \checkmark	 & \checkmark	& \\
 4654	 & 1987	 & \checkmark	 & \checkmark	 &  	 &  	 & \checkmark	& a\\
 4689	 & 2058	 &  	 &  	 &  	 &  	 &  	& \\
\dots	 	 & 2062	 &  	 &  	 & \checkmark	 &  	 &  	& \\
 4694	 & 2066	 &  	 &  	 & \checkmark	 &  	 &  	& \\
 4698	 & 2070	 &  	 & \checkmark	 & \checkmark	 &  	 & \checkmark	& \\
 4713	 & \dots	 &  	 &  	 &  	 & \checkmark	 &  	& \\
 4772	 & \dots	 &  	 &  	 & \checkmark	 & \checkmark	 &  	& \\
 4808	 & \dots	 &  	 &  	 & \checkmark	 &  	 &  	& \\		
\hline	
\end{tabular}
    \begin{tablenotes}
     \item The evaluations from Chung et al. (2009); whether the \hi \ shows evidence of stripping, ram-pressure stripping, tidal interaction, a warp or an \hi \ tail.
     \item [a] Identified in \cite{Chung07} as a galaxy with a one-sided \hi \ tail. 
     \item [I-V] Order in the time sequence of ram-pressure stripping according to \protect\cite{Vollmer09}.
     \end{tablenotes}
\label{t:chung}
\end{threeparttable}%

\newpage



\begin{table*}
\caption{ The morphological values and errors of the VIVA \hi \ column density maps.}
\begin{tabular}{l l l l l l l l}
Galaxy  		& Gini   & $M_{20}$   & $C_{20/80}$   & Asymmetry   & Smoothness      & $G_M$   \\ 
\hline
\hline
NGC 4064 	 & 0.342 $\pm$ 0.041 	 & -0.652 $\pm$ 0.195 	 & 0.810 $\pm$ 0.430 	 & 2.000 $\pm$ 0.000 	 & 0.501 $\pm$   \dots 	 & 0.206 $\pm$   \dots \\ 
NGC 4189 	 & 0.346 $\pm$ 0.011 	 & -0.691 $\pm$ 0.019 	 & 1.156 $\pm$ 0.106 	 & 2.000 $\pm$ 0.000 	 & 0.154 $\pm$ 0.030 	 & 0.266 $\pm$ 0.077 \\ 
NGC 4192 	 & 0.444 $\pm$ 0.008 	 & -0.965 $\pm$ 0.199 	 & 7.162 $\pm$ 2.317 	 & 1.171 $\pm$ 0.434 	 & 0.200 $\pm$ 0.022 	 & 0.687 $\pm$ 0.033 \\ 
NGC 4216 	 & 0.398 $\pm$ 0.012 	 & -0.633 $\pm$ 0.026 	 & 0.993 $\pm$ 0.284 	 & 2.000 $\pm$ 0.000 	 & 0.241 $\pm$ 0.024 	 & 0.820 $\pm$ 0.132 \\ 
NGC 4222 	 & 0.000 $\pm$ 0.000 	 & 0.000 $\pm$ 0.000 	 & 0.000 $\pm$ 0.000 	 & 0.000 $\pm$ 0.000 	 & 0.000 $\pm$ 0.000 	 & 0.000 $\pm$ 0.000 \\ 
NGC 4254 	 & 0.451 $\pm$ 0.004 	 & -0.666 $\pm$ 0.008 	 & 1.580 $\pm$ 0.075 	 & 2.000 $\pm$ 0.000 	 & 0.035 $\pm$ 0.012 	 & 0.232 $\pm$ 0.022 \\ 
NGC 4293 	 & 0.114 $\pm$   \dots 	 & -0.758 $\pm$ 0.089 	 & 1.933 $\pm$ 2.181 	 & 1.605 $\pm$   \dots 	 & 1.043 $\pm$   \dots 	 & 0.780 $\pm$   \dots \\ 
NGC 4294 	 & 0.493 $\pm$ 0.025 	 & -0.730 $\pm$ 0.050 	 & 2.310 $\pm$ 0.462 	 & 2.000 $\pm$ 0.000 	 & 0.280 $\pm$ 0.038 	 & 0.777 $\pm$ 0.064 \\ 
NGC 4298 	 & 0.525 $\pm$ 0.009 	 & -0.631 $\pm$ 0.017 	 & 1.749 $\pm$ 0.246 	 & 2.000 $\pm$ 0.000 	 & 0.229 $\pm$ 0.034 	 & 0.178 $\pm$ 0.080 \\ 
NGC 4299 	 & 0.175 $\pm$   \dots 	 & -0.683 $\pm$ 0.136 	 & 2.299 $\pm$ 1.288 	 & 2.000 $\pm$ 0.000 	 & 0.373 $\pm$   \dots 	 & 0.279 $\pm$   \dots \\ 
NGC 4302 	 & 0.520 $\pm$ 0.008 	 & -0.976 $\pm$ 0.129 	 & 7.231 $\pm$ 1.485 	 & 1.932 $\pm$ 0.487 	 & 0.228 $\pm$ 0.040 	 & 0.259 $\pm$ 0.061 \\ 
NGC 4321 	 & 0.301 $\pm$ 0.008 	 & -0.769 $\pm$ 0.018 	 & 1.994 $\pm$ 0.579 	 & 2.000 $\pm$ 0.000 	 & 0.127 $\pm$ 0.016 	 & 0.138 $\pm$ 0.106 \\ 
NGC 4330 	 & 0.474 $\pm$ 0.018 	 & -0.688 $\pm$ 0.110 	 & 0.362 $\pm$ 0.105 	 & 2.000 $\pm$ 0.000 	 & 0.499 $\pm$ 0.044 	 & 0.661 $\pm$ 0.151 \\ 
NGC 4351 	 & 0.397 $\pm$ 0.013 	 & -0.693 $\pm$ 0.027 	 & 1.916 $\pm$ 0.393 	 & 2.000 $\pm$ 0.000 	 & 0.148 $\pm$ 0.030 	 & 0.249 $\pm$ 0.107 \\ 
NGC 4380 	 & 0.210 $\pm$ 0.009 	 & -0.799 $\pm$ 0.061 	 & 1.725 $\pm$ 0.676 	 & 2.000 $\pm$ 0.000 	 & 0.293 $\pm$ 0.024 	 & 0.551 $\pm$ 0.151 \\ 
NGC 4383 	 & 0.306 $\pm$ 0.013 	 & -0.709 $\pm$ 0.032 	 & 2.442 $\pm$ 0.566 	 & 2.000 $\pm$ 0.000 	 & 0.127 $\pm$ 0.023 	 & 0.149 $\pm$ 0.116 \\ 
NGC 4388 	 & 0.525 $\pm$ 0.020 	 & -0.799 $\pm$ 0.201 	 & 2.957 $\pm$ 0.878 	 & 1.654 $\pm$ 0.262 	 & 0.386 $\pm$ 0.089 	 & 0.546 $\pm$ 0.082 \\ 
NGC 4394 	 & 0.188 $\pm$ 0.005 	 & -0.764 $\pm$ 0.081 	 & 2.591 $\pm$ 0.503 	 & 2.000 $\pm$ 0.000 	 & 0.195 $\pm$ 0.008 	 & 0.094 $\pm$ 0.070 \\ 
NGC 4396 	 & 0.482 $\pm$ 0.022 	 & -0.686 $\pm$ 0.049 	 & 1.753 $\pm$ 0.446 	 & 2.000 $\pm$ 0.000 	 & 0.400 $\pm$ 0.024 	 & 0.700 $\pm$ 0.150 \\ 
NGC 4405 	 & 0.314 $\pm$ 0.031 	 & -0.666 $\pm$ 0.238 	 & 0.872 $\pm$ 0.609 	 & 2.000 $\pm$ 0.000 	 & 0.445 $\pm$   \dots 	 & 0.203 $\pm$   \dots \\ 
NGC 4402 	 & 0.000 $\pm$ 0.000 	 & 0.000 $\pm$ 0.000 	 & 0.000 $\pm$ 0.000 	 & 0.000 $\pm$ 0.000 	 & 0.000 $\pm$ 0.000 	 & 0.000 $\pm$ 0.000 \\ 
IC 3355 	 	 & 0.000 $\pm$ 0.000 	 & 0.000 $\pm$ 0.000 	 & 0.000 $\pm$ 0.000 	 & 0.000 $\pm$ 0.000 	 & 0.000 $\pm$ 0.000 	 & 0.000 $\pm$ 0.000 \\ 
NGC 4419 	 & 0.328 $\pm$   \dots 	 & -0.661 $\pm$ 0.184 	 & 0.907 $\pm$ 0.433 	 & 2.000 $\pm$   \dots 	 & 0.657 $\pm$ 0.044 	 & 0.773 $\pm$ 0.182 \\ 
NGC 4424 	 & 0.484 $\pm$ 0.021 	 & -0.674 $\pm$ 0.042 	 & 1.074 $\pm$ 0.302 	 & 2.000 $\pm$ 0.000 	 & 0.328 $\pm$ 0.071 	 & 0.496 $\pm$ 0.141 \\ 
NGC 4450 	 & 0.000 $\pm$ 0.000 	 & 0.000 $\pm$ 0.000 	 & 0.000 $\pm$ 0.000 	 & 0.000 $\pm$ 0.000 	 & 0.000 $\pm$ 0.000 	 & 0.000 $\pm$ 0.000 \\ 
IC 3392 	 	 & 0.320 $\pm$ 0.029 	 & -0.685 $\pm$ 0.263 	 & 0.367 $\pm$ 0.219 	 & 2.000 $\pm$ 0.000 	 & 0.434 $\pm$   \dots 	 & 0.458 $\pm$   \dots \\ 
NGC 4457 	 & 0.287 $\pm$ 0.016 	 & -0.716 $\pm$ 0.033 	 & 0.908 $\pm$ 0.216 	 & 2.000 $\pm$ 0.000 	 & 0.234 $\pm$ 0.034 	 & 0.108 $\pm$ 0.149 \\ 
NGC 4501 	 & 0.331 $\pm$ 0.007 	 & -0.603 $\pm$ 0.024 	 & 2.201 $\pm$ 1.120 	 & 0.882 $\pm$ 0.547 	 & 0.185 $\pm$ 0.016 	 & 0.501 $\pm$ 0.035 \\ 
NGC 4522 	 & 0.398 $\pm$ 0.018 	 & -0.660 $\pm$ 0.027 	 & 0.680 $\pm$ 0.336 	 & 2.000 $\pm$ 0.000 	 & 0.251 $\pm$ 0.045 	 & 0.478 $\pm$ 0.078 \\ 
NGC 4532 	 & 0.591 $\pm$ 0.008 	 & -0.693 $\pm$ 0.010 	 & 1.086 $\pm$ 0.079 	 & 2.000 $\pm$ 0.000 	 & 0.112 $\pm$ 0.056 	 & 0.256 $\pm$ 0.088 \\ 
NGC 4535 	 & 0.322 $\pm$ 0.007 	 & -0.736 $\pm$ 0.048 	 & 2.303 $\pm$ 0.341 	 & 2.000 $\pm$ 0.000 	 & 0.156 $\pm$ 0.014 	 & 0.210 $\pm$ 0.068 \\ 
NGC 4533 	 & 0.000 $\pm$ 0.000 	 & 0.000 $\pm$ 0.000 	 & 0.000 $\pm$ 0.000 	 & 0.000 $\pm$ 0.000 	 & 0.000 $\pm$ 0.000 	 & 0.000 $\pm$ 0.000 \\ 
NGC 4536 	 & 0.371 $\pm$ 0.005 	 & -0.734 $\pm$ 0.034 	 & 2.771 $\pm$ 0.855 	 & 0.407 $\pm$ 0.451 	 & 0.097 $\pm$ 0.025 	 & 0.562 $\pm$ 0.034 \\ 
NGC 4548 	 & 0.211 $\pm$ 0.004 	 & -0.808 $\pm$ 0.116 	 & 4.793 $\pm$ 1.165 	 & 0.518 $\pm$ 0.454 	 & 0.214 $\pm$ 0.013 	 & 0.223 $\pm$ 0.066 \\ 
NGC 4561 	 & 0.417 $\pm$ 0.008 	 & -0.694 $\pm$ 0.009 	 & 2.407 $\pm$ 0.366 	 & 2.000 $\pm$ 0.000 	 & 0.203 $\pm$ 0.032 	 & 0.395 $\pm$ 0.049 \\ 
NGC 4567 	 & 0.424 $\pm$ 0.010 	 & -0.710 $\pm$ 0.018 	 & 1.152 $\pm$ 0.182 	 & 2.000 $\pm$ 0.000 	 & 0.115 $\pm$ 0.024 	 & 0.439 $\pm$ 0.032 \\ 
NGC 4568 	 & 0.424 $\pm$ 0.008 	 & -1.096 $\pm$ 0.199 	 & 5.611 $\pm$ 1.421 	 & 2.000 $\pm$ 0.321 	 & 0.115 $\pm$ 0.028 	 & 0.439 $\pm$ 0.044 \\ 
NGC 4569 	 & 0.434 $\pm$ 0.010 	 & -0.703 $\pm$ 0.019 	 & 0.429 $\pm$ 0.079 	 & 1.282 $\pm$ 0.369 	 & 0.213 $\pm$ 0.021 	 & 0.519 $\pm$ 0.200 \\ 
NGC 4579 	 & 0.218 $\pm$ 0.013 	 & -0.705 $\pm$ 0.050 	 & 1.199 $\pm$ 0.330 	 & 2.000 $\pm$ 0.000 	 & 0.203 $\pm$ 0.026 	 & 0.240 $\pm$ 0.174 \\ 
NGC 4580 	 & 0.201 $\pm$   \dots 	 & -0.674 $\pm$ 0.200 	 & 0.694 $\pm$ 0.386 	 & 2.000 $\pm$   \dots 	 & 0.469 $\pm$ 0.032 	 & 0.364 $\pm$   \dots \\ 
NGC 4606 	 & 0.467 $\pm$ 0.018 	 & -0.675 $\pm$ 0.039 	 & 0.108 $\pm$ 0.032 	 & 2.000 $\pm$ 0.000 	 & 0.338 $\pm$ 0.067 	 & 0.669 $\pm$ 0.115 \\ 
NGC 4607 	 & 0.463 $\pm$ 0.019 	 & -0.667 $\pm$ 0.066 	 & 0.778 $\pm$ 3.209 	 & 2.000 $\pm$ 0.000 	 & 0.337 $\pm$ 0.057 	 & 0.645 $\pm$ 0.073 \\ 
NGC 4651 	 & 0.360 $\pm$ 0.006 	 & -0.712 $\pm$ 0.008 	 & 1.597 $\pm$ 0.105 	 & 2.000 $\pm$ 0.000 	 & 0.113 $\pm$ 0.024 	 & 0.590 $\pm$ 0.022 \\ 
NGC 4654 	 & 0.442 $\pm$ 0.007 	 & -1.279 $\pm$ 0.258 	 & 8.395 $\pm$ 1.295 	 & 1.097 $\pm$ 0.502 	 & 0.120 $\pm$ 0.021 	 & 0.522 $\pm$ 0.043 \\ 
NGC 4689 	 & 0.249 $\pm$ 0.010 	 & -0.739 $\pm$ 0.074 	 & 2.942 $\pm$ 1.516 	 & 2.000 $\pm$ 0.000 	 & 0.126 $\pm$ 0.010 	 & 0.173 $\pm$ 0.065 \\ 
VCC 2062 	 & 0.000 $\pm$ 0.000 	 & 0.000 $\pm$ 0.000 	 & 0.000 $\pm$ 0.000 	 & 0.000 $\pm$ 0.000 	 & 0.000 $\pm$ 0.000 	 & 0.000 $\pm$ 0.000 \\ 
NGC 4694 	 & 0.297 $\pm$ 0.011 	 & -0.693 $\pm$ 0.037 	 & 0.966 $\pm$ 0.501 	 & 2.000 $\pm$ 0.000 	 & 0.377 $\pm$ 0.054 	 & 0.581 $\pm$ 0.099 \\ 
NGC 4698 	 & 0.227 $\pm$ 0.003 	 & -0.739 $\pm$ 0.014 	 & 3.115 $\pm$ 0.312 	 & 2.000 $\pm$ 0.000 	 & 0.145 $\pm$ 0.008 	 & 0.583 $\pm$ 0.021 \\ 
NGC 4713 	 & 0.398 $\pm$ 0.005 	 & -0.695 $\pm$ 0.008 	 & 2.082 $\pm$ 0.149 	 & 2.000 $\pm$ 0.000 	 & 0.110 $\pm$ 0.021 	 & 0.202 $\pm$ 0.045 \\ 
NGC 4772 	 & 0.460 $\pm$ 0.007 	 & -0.734 $\pm$ 0.061 	 & 2.733 $\pm$ 0.424 	 & 2.000 $\pm$ 0.032 	 & 0.329 $\pm$ 0.049 	 & 0.511 $\pm$ 0.075 \\ 
NGC 4808 	 & 0.498 $\pm$ 0.010 	 & -0.696 $\pm$ 0.016 	 & 2.883 $\pm$ 0.481 	 & 2.000 $\pm$ 0.000 	 & 0.151 $\pm$ 0.049 	 & 0.417 $\pm$ 0.111 \\

\hline	
\end{tabular}
\label{t:viva}
\end{table*}%

\end{document}